\documentclass[12pt]{article}
\usepackage{graphicx}
\usepackage{subfigure}
\begin{document}
\begin{titlepage}
\begin{center}
{\Large {\bf A NEW APPROACH TO DYNAMIC FINITE-SIZE SCALING}}
\end{center}

\vskip 1.5cm

\centerline{MEHMET D\.ILAVER , SEMRA G\"UND\"U\c{C},  
MERAL AYDIN$^{\dagger}$ and Y\.I\u{G}\.IT G\"UND\"U\c{C}}

\centerline{\it Hacettepe University, Physics Department,}
\centerline{\it 06532  Beytepe, Ankara, Turkey }
\centerline{\it  $\dagger$ \c{C}ankaya University, Mathematics and Computer Science Department, \\}
\centerline{\it 06530 Balgat, Ankara, Turkey}

\vskip 0.5cm

\centerline{\normalsize {\bf Abstract} }

{\small
In this work we have considered the Taylor series expansion of the
dynamic scaling relation of the magnetization with respect to small initial
magnetization values in order to study the dynamic scaling behaviour of
$2$- and $3$-dimensional Ising models.  We have used the literature
values of the critical exponents and of the new dynamic exponent $x_0$ to
observe the dynamic finite-size scaling behaviour of the time
evolution of the magnetization during early stages of the Monte Carlo
simulation. For $3$-dimensional Ising Model we have also presented that this method opens 
the possibility of calculating $z$ and $x_0$ separately. Our results show good agreement with the 
literature values.  Measurements done on lattices with different sizes  seem to give
very good scaling. }

\vskip 0.5cm

{\small {\it Keywords:} Ising Model, Dynamic Scaling, Time 
Evolution of the Magnetization }

\end{titlepage}        

\pagebreak

\section{Introduction}

The determination of the critical temperature and the corresponding
critical exponents for the systems exhibiting second-order phase
transitions is the major task in many areas of physics. In order to
determine the critical parameters of the system there exist two
approaches. Traditionally, one uses universality and the scaling
relations in the equilibrium stages of the statistical mechanical
system. In other words, one studies the long-time regime of the
system.  The second approach is to study dynamic scaling which seems
to exist in the early stages of the quenching process in the system
exhibiting second-order phase transition.  Jansen, Schaub and
Schmittmann~\cite{Janss89} have shown that for a dynamic relaxation
process, in which a system is evolving according to a dynamics of
Model A~\cite{Hoh77} and is quenched from a very high temperature to
the critical temperature, a universal dynamic scaling behaviour within
the short-time regime exists~\cite{Zheng98,Zheng00}.

$\;$

Monte Carlo simulations are the most commonly used tools to study
statistical mechanical systems in thermal equilibrium around the phase
transition point. Creating statistically independent configurations
is controlled by the autocorrelations in the system which depend on
the system as well as the simulation algorithm. Around the phase
transition point, the correlation length grows exponentially and this
results in the effect known as the critical slowing down. On finite
lattices the autocorrelation time diverges as $\tau \sim  L^z$, where $L$
is the linear size of the lattice and $z$ is the dynamic critical
exponent.

$\;$

Dynamic finite-size scaling relation is a generalized form of the
finite-size scaling relation valid for the equilibrium stages. 
For the $k^{th}$ moment of the magnetization of a system with spatial
size $L$ this relation can be written as~\cite{Janss89}

\begin{equation}
\label{magnetization}
M^{(k)}(t,\epsilon,{m_0,L})=L^{(-k{\beta}/{\nu})}{\cal{M}}^{(k)}(tL^{-z},{\epsilon}L^{1/{\nu}},{m_{0}}L^{x_0})
\end{equation}

where $\beta$ and $\nu$ are the well-known critical exponents, $t$ is the
simulation time and ${\epsilon}=(T-{T_c})/{T_c}$ is the reduced
temperature. In the study of the short-time dynamic scaling of a
system, in addition to the critical exponents, the dynamic critical
exponent $z$ and a new and independent exponent $x_0$ appear, where
$x_0$ is the anomalous dimension of the initial magnetization $m_0$.

$\;$

Eq.(\ref{magnetization}) can be reformulated for the time evolution of
the magnetization at early stages of the simulation. For very small
initial magnetization, time evolution of the magnetization undergoes
a critical initial increase~\cite{Janss89,Zheng98} in the form

\begin{equation}
\label{mlike}
M(t) \sim {m_{0}}{t^{\theta}}
\end{equation}

where the exponent $\theta$ is related to the exponent $x_0$ by
$\theta=({x_0}-{\beta}/{\nu})/z$. The power-law behaviour
relation (Eq.(\ref{mlike})) of the dynamic finite-size scaling is used
to estimate the critical exponents $z$, $\theta$ and ${\beta}/{\nu}$. 
In  literature, it has been shown that the dynamic behaviour of
various spin models obeys the dynamic scaling relations
(Eqs. (\ref{magnetization}) and (\ref{mlike}))
~\cite{Ito93,Schue95,Okano97,Ying01,Luo99,Zheng99,Jaster99,Okano98}; the
universality of $x_0$ has also been studied 
and numerical evidence for the existence of the universality of the
dynamic scaling has been given~\cite{Okano97,Zhang99}. 

$\;$

In this work we aim to reconsider
dynamic finite-size scaling by approaching the problem from a
different point of view. Instead of the critical initial increase \mbox{(Eq. (\ref{mlike}))} of
the magnetization used in the literature, we have used the
dynamic scaling relation (Eq.(\ref{magnetization})) by considering the
Taylor series expansion of ${{M}}^{(k)}(t,{\epsilon},m_0,L)$ with 
respect to small initial magnetization $m_0$. We have seen that the new exponent ($x_0$)
exists not only in the initial stages, but also until the effects of the initial 
magnetization of the system diminish. This approach seems to
give the critical exponents and the dynamic exponents with very high
accuracy. In order to discuss the proposed method we have considered
$2$- and $3$-dimensional Ising models which are well-studied by finite-size scaling
 techniques for both  equilibrium and non-equilibruium stages.

$\;$

In the next section we present the model, the method and the scaling relation
for the magnetization. In section 3 numerical
results for $2$- and $3$-dimensional Ising models will be reported. The
last section contains the conclusions.

\section{Model and the Method}

In this work we have employed $2$- and $3$-dimensional Ising models which
are described by the Hamiltonian

\begin{equation}
\label{hamiltonian}
{-\beta}H = K{\sum_{<ij>}}{S_{i}}{S_{j}} \;.
\end{equation}

Here $\beta=1/kT$ and $K=J/kT$, where $k$ is the Boltzmann constant,
$T$ is the temperature and $J$ is the magnetic interaction between the
spins. In the Ising Model the spin variables take the values
${S_{i}}={\pm 1}$. In this work we have performed Monte Carlo
simulations by using Metropolis algorithm.

$\;$

The measurements of the thermodynamic quantities which are obtained
using thermalized configurations in Monte Carlo simulations are
independent of the initial configuration. Here, before the
thermalization, the convergence path (to the thermalization) may be
different depending on which initial configuration is used. Although
the system reaches the same thermalized configurations in the \mbox{long-time} 
regime, the convergence path is highly sensitive to the initial
configuration. In this work we propose to consider the dynamics of the
thermalization for a system with two different initial configurations,
in order to obtain the new dynamic index $x_0$. One initial
configuration can be considered as the configuration with vanishing
initial magnetization. The second initial configuration can be taken
with arbitrary, but small initial magnetization. The dynamic
finite-size scaling form of the $k^{th}$ moment of the magnetization
is given by Eq.(\ref{magnetization}). At the critical point
$({\epsilon}=0)$, for small initial magnetization ($m_0$) values, the
function given in Eq.(\ref{magnetization}) can be expanded into Taylor
series around $m_0=0$ resulting in

\begin{eqnarray}
\label{taylor}
M^{(k)}(t,0,{m_0},L) &=& L^{-k{\beta}/{\nu}}{\cal{M}}^{(k)}(tL^{-z},0,0) +\\ \nonumber
&&\frac{\partial{\cal{M}}^{(k)}(tL^{-z},0,{m_0})}
{\partial{m_{0}}}{\mid}_{ {m_{0}}=0}{m_0}L^{{x_0}-{k\beta}/{\nu}}+O({m_0}^2) \;.
\end{eqnarray}

The first term in this expansion has no $m_0$ dependence and it can be
calculated through Monte Carlo simulations by considering the initial
configuration with vanishing initial magnetization. The
$m_0$-dependent terms (second and higher order terms in the expansion)
can be represented by a function $M_s(t,0,m_0,L)$ which is basically

\begin{equation}
\label{MS}
M_s(t,0,m_0,L)\,=\,\frac{\partial{\cal{M}}^{(k)}(tL^{-z},0,{m_0})}
{\partial{m_{0}}}{\mid}_{ {m_{0}}=0}{m_0}L^{{x_0}-{k\beta}/{\nu}}+O({m_0}^2) \;
\end{equation}

and it can be obtained by calculating the difference of $M^{(k)}(t,0,m_0,L)$
and $M^{(k)}(t,0,0,L)$. This means it has the form,

\begin{equation}
\label{diff}
M_s(t,0,m_0,L)=M^{(k)}(t,0,m_0,L)-M^{(k)}(t,0,0,L).   
\end{equation}

Here $M^{(k)}(t,0,0,L)$ corresponds to the first term in the
expansion. 
The terms $M^{(k)}(t,0,m_0,L)$ and $M^{(k)}(t,0,0,L)$ can be measured
through Monte Carlo simulations using two different initial
configurations, one with initial magnetization $m_0$ and the other one
with vanishing initial magnetization.

$\;$

The function given in Eqs.(\ref{MS}) and (\ref{diff}) contains all $m_0$ dependence.
The thermalized configurations are independent of the initial
magnetization, hence $M_s(t,0,m_0,L)$ is a decaying function.
This function is expected to be a function of the autocorrelation time 
and the initial magnetization. In this work, we have chosen
an exponentially decaying function in the form

\begin{equation}
\label{phi2}
M_s(t,0,m_0,L)= f(m_0,L)\;e^{{-(t/\tau(L))}^{\phi(L)}}.
\end{equation}

Here $t$ is the simulation time, $\tau(L)$ is the autocorrelation time
and $\phi(L)$ is an exponent which is related to the initial slip. If this form
fits the data, two dynamical exponents can be calculated separately.
$f(m_0,L)$ is a $m_0$-dependent function, but it is has no autocorrelation 
time dependence. Hence  the new dynamic
scaling exponent $x_0$ can be calculated from the size dependence of
this function for small values of $m_0$. The autocorrelation time and hence $z$
 can be calculated from the decay rate of $M_s(t,0,m_0,L)$. The validity of our
assumptions has been tested by fitting the function given in
Eq(\ref{phi2}) to our data and the results are presented in Section 3.

$\;$

For very small initial magnetization values, it is expected that the
leading term in the expansion (Eq. (\ref{taylor})) is dominant and the
rest of the terms (represented by the function $M_s(t,0,m_0,L)$) is
in the same order of magnitude with the simulation errors.  Increasing
initial magnetization results in a sizeable $m_0$-dependent term in
Eq. (\ref{taylor}). In this range of $m_0$ one can expect that
$M_s(t,0,m_0,L)$ scales like $L^{x_0 - \beta / \nu}$. Further increase
in $m_0$ reduces the reliability of such scaling since the
contribution of higher order terms in the expansion
(Eq. (\ref{taylor})) becomes effective. In our simulations we have
used different $m_0$ values to investigate the proposed scaling
behaviour of $M_s(t,0,m_0,L)$. To test the scaling we have used the
literature values of $\beta/{\nu}$ and $x_0$ for $2$- and
$3$-dimensional Ising models and also calculated the dynamic exponents
$z$ and $x_0$ for $3$-dimensional model through a fit to our Monte
Carlo data.

\section{Results and Discussions}

We have studied $2$-and $3$-dimensional Ising models evolving in time
according to the dynamics of the Model A\cite{Hoh77}. We have
prepared the lattices with varying initial magnetization values
starting from $m_{0}=0$. Totally random initial configurations are
quenched at the corresponding infinite lattice critical
temperature. Simulations are performed by using Metropolis algorithm
on lattices with linear sizes $L= 512, 640, 768, 896$ and $L=32, 64,
96$ for $2$- and $3$-dimensional Ising models respectively. For error
calculations we have binned data points obtained from $4000$ to $15000$ independent runs for 
$3$-dimensional model,
and $50000$ to $500000$ independent runs for $2$-dimensional model, starting from 
randomly prepared configurations.  For
$3$-dimensional model, $3000$ iterations are performed for each
independent run. For $2$-dimensional model there are two different
sets of runs. One set of data is obtained after $500$ iterations for
each independent run and for the other set of data, each independent
run includes $10000$ iterations.

$\;$

The dynamic critical exponent $z$ has a little dependence on
dimension for local algorithms. 
For a given numbers of spins, the saturation in magnetization can be observed
after relatively small number of iterations in $3$-dimensional lattice compared two
$2$-dimensional one, since the saturation is observed when the correlation length reaches
the linear size of the lattice.  
In order to observe the behaviour of the $m_0$-dependent terms, one needs to study
the time development of the system until both $M^{(k)}(t,0,m_0,L)$ and
$M^{(k)}(t,0,0,L)$ reach approximately the same value.  In our
simulation studies, we have observed that for $3$-dimensional lattice,
configurations with both vanishing and non-vanishing initial
magnetization reach the same value in (approximately) $3000$
iterations for the largest lattice $(L=96)$. For $2$-dimensional
model, since much larger linear sizes are used, the initial
configuration dependence remains after $30000$ iterations and
observing the decaying behaviour of $M_s(t,0,m_0,L)$ becomes extremely
expensive. For this reason, we have tested our ideas for
$3$-dimensional model. In $2$-dimensional lattice, we have performed
simulations up to $10000$ iterations, but even this many iterations is
extremely short to observe the decaying behaviour of $M_s(t,0,m_0,L)$.
Nevertheless, it is sufficient to test the scaling behaviour and to
obtain the new dynamic critical exponent $x_0$.

$\;$

In Figure 1, the magnetization data are presented for vanishing
initial magnetization and for $m_0 \approx 0.1$ in $3$-dimensional
Ising Model with linear sizes $L= 32, 64, 96$. The magnetization data
converge approximately to the same value for both initial
configurations after approximately $500 (L=32), 1500 (L=64), 3000
(L=96)$ iterations. The differences between the data obtained by
starting from different initial magnetizations are large initially for
large lattices and vanishes after above mentioned number of iterations
are reached. The final magnetization value decreases  as the lattice
size increases, and the magnetization is smaller initially for
vanishing $m_0$. The function $M_s(t,0,m_0,L)$ which represents the
difference between $M^{(k)}(t,0,m_0,L)$ and $M^{(k)}(t,0,0,L)$ is
given in Figure 2, for $m_0 \approx 0.1$ and  for linear sizes, $L=32, 64,96$ in
$3$-dimensional lattice. The function $M_s(t,0,m_0,L)$ is a decreasing
function with an initial value proportional to the lattice size
$L$. Despite the fact that our main aim is not to obtain the dynamical
critical exponent $(z)$, since we are interested in the functional
form of the $m_0$-dependent term and the speed of its decay, we have
fitted our data to the functional form $(\tau \sim L^z)$ and obtained
$z$ from the fit results.  We have fitted the data to the exponential
function given in Eq.(\ref{phi2}).  This function fits the data
reasonably well, as it can be seen from Figure 2. The correlation times
$(\tau)$ obtained from the fit are $\tau(L) = 80,\; 372,\;
839$ for $L = 32, 64, 96$ respectively. The errors are in the order of one.
We have fitted the functional form $(\tau \sim L^z)$ to our data and we 
have obtained the value for $z$ as $z=2.04\pm0.07$ for $3$-dimensional model.
Figure 3 shows the fitted function and the data points for three
lattice sizes used.

$\;$

As it is mentioned in Section 2, the function $M_s(t,0,m_0,L)$
representing $m_0$-dependent terms scales like $L^{z\theta}$.
Here we have used the literature values of the exponent
$\theta=(x_0-\beta/\nu)/z$, $\theta=0.191(2)$~\cite{Zhang99} and
$\theta=0.108(2)$~\cite{Jaster99} for $2$- and $3$-dimensional Ising
models respectively, to test the scaling.  The exponent $z\,\theta = x_0-\beta/\nu$
is obtained from these literature values of both $\theta$ and $z$.
For $2$-dimensional Ising Model, the value of the dynamic critical
exponent $z$ for Metropolis algorithm varies from $2.13$ to $2.18$
~\cite{Zheng98,Zheng00,Ito93,Tang87,Landau88,Li95,Li96,Grassberger95,Nightingale96,Gropengiesser}. 
For the $3$-dimensional case, the equilibrium case, the value of the
dynamic critical exponent is $z=2.04(3)$~\cite{Wansleben91}, while
from the dynamic scaling, $z$ ranges from $2.032(11)$ to
$2.052(17)$~\cite{Jaster99}.  We have used $z=2.18 $ and $z=2.04$ for
$2$- and $3$-dimensional Ising models respectively. The scaling of the function  
$M_s(t,0,m_0,L)$ obtained from the Monte Carlo data
for $3$-dimensional model is given in Figure 4. As it can be seen from this figure, these
terms are scaled reasonably well.  We have also
fitted the data to a function to obtain the best fit and to obtain the
value for $z\theta$. Our value $z\,\theta=0.27{\pm}0.03$ (for $3$-dimensional model) is 
slightly larger than the value in the literature.

$\;$ 

For $2$-dimensional Ising Model, the autocorrelation time $(\tau)$ is
extremely long for the lattice sizes we use. Consequently, we didn't
aim to see the decaying form of $M_s(t,0,m_0,L)$. Instead we have
tested the scaling behaviour of $M_s(t,0,m_0,L)$.  In Figure 5, we have
presented the data for $M_s(t,0,m_0,L)$ for after $500$ iterations and
Figure 6 shows the scaling of the  data when the literature values are
used.

\section{Conclusions}

The short-time dynamic behaviour of the magnetization is studied in
$2$- and $3$-dimensional Ising models.  The time-dependent
magnetization profiles with different initial magnetization ($m_0$)
values exhibit an initial increase during the time scale set by the
initial magnetization, the dynamic critical exponent ($z$) and the new
dynamic exponent ($x_0$). After a long-time ($t\sim L^z$), all profiles 
gradually join together, since in the
long-time regime the effects of the initial conditions are ignorable.
In this work we aimed to study the effects of the initial
configurations on magnetization by using Monte Carlo simulation. This can be
observed by expanding the magnetization around $m_0=0$. This expansion 
(Eq. (\ref{magnetization}))  gives a leading term which is
independent of $m_0$ and the terms containing the information of how
the initial configuration is prepared. 

$\;$

In this approach, the dynamic critical behaviour can be studied by
using the data from the early stages of the Monte Carlo simulation. In
order to obtain a relation between the size and the initial
magnetization dependence of the Monte Carlo data, we have considered
both vanishing and non-vanishing initial magnetization runs. The
iteration by iteration differences of these two sets of runs yield the
initial magnetization dependence of the system. Using different
lattice sizes we have shown that the proposed scaling holds.  We have
also shown that the dynamic finite-size scaling is valid not only for
the very early stages (within $100$ iterations) of the Monte Carlo
runs, but it is valid until the effects of the the initial
magnetization diminish.

$\;$

In order to test our proposed approach, we have employed the literature
values of $\beta/\nu$, $z$, $x_0$ and we have observed a good scaling
behaviour for $m_0$-dependent terms for the lattice sizes we  used. Such a good
scaling gives hopes for the applicability of this approach to a wide range
of statistical mechanical systems for calculations of both dynamic and
equilibrium critical exponents.

\pagebreak

\section*{Acknowledgements}

We greatfully acknowledge Hacettepe University Research Fund 
\mbox{(Project no : 01 01 602 019)} and Hewlett-Packard's Philanthropy Programme.

\pagebreak

\pagebreak

\section*{Figure Captions}

Figure 1.    Magnetization data for $3$-dimensional Ising Model with
linear lattice sizes $L=32,64,96$, with vanishing and nonzero initial magnetization
($m_0 \approx 0.1$) for each size, as a function of simulation time $t$. 
The curves for vanishing initial magnetization start from the origin.

$\;$

Figure 2.   The function $M_s (t,0,m_0,L)$ corresponding to $m_{0}$-dependent
terms in the expansion as a function of simulation time $t$, for $3$-dimensional
Ising Model. The plot of the function fitted to the form given in Eq.(\ref{phi2}) is also shown.

$\;$ 

Figure 3. The autocorrelation time $\tau(L)$ as a function of the linear lattice size
$L = 32,64,96$ in $3$-dimensional Ising Model. The function fitted to the form 
$\tau \sim L^z$ is also shown.

$\;$ 

Figure 4. Scaled form of  $M_s (t,0,m_0,L)$ obtained from the simulation data (given 
in Figure 2), for $3$-dimensional Ising Model.

$\;$

Figure 5.  $M_s (t,0,m_0,L)$ obtained from the simulation data as a function of simulation time
$t$ for $2$-dimensional Ising Model ($m_0 \approx 0.07$).

$\;$

Figure 6. Scaled form of the data given in Figure 5 for function $M_s (t,0,m_0,L)$ for 
$2$-dimensional Ising Model.

\pagebreak

\begin{figure}
\centering
\includegraphics[angle=0,width=10truecm]{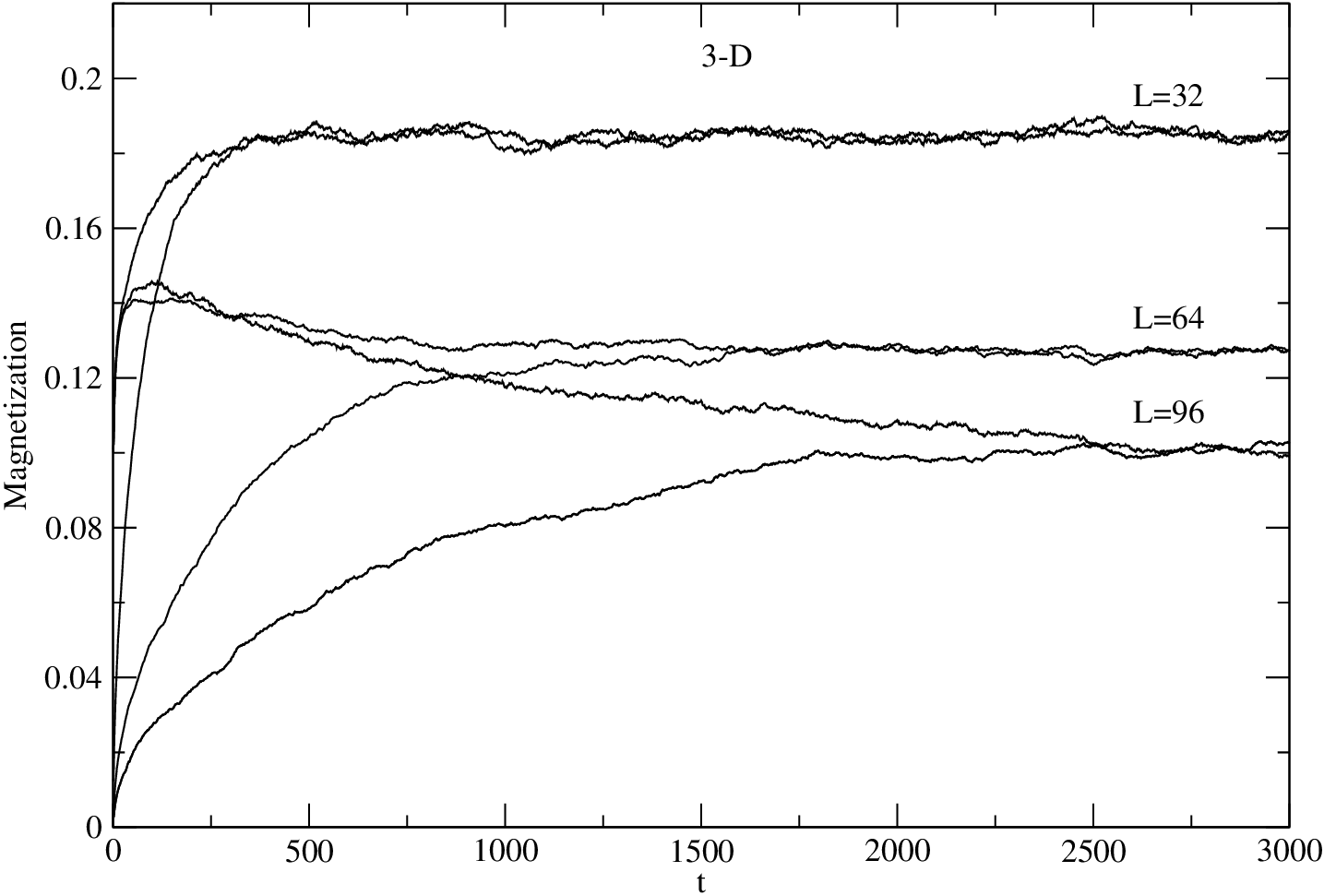}
\caption{}
\label{tttt1}
\end{figure}

\begin{figure}
\centering
\includegraphics[angle=0,width=10truecm]{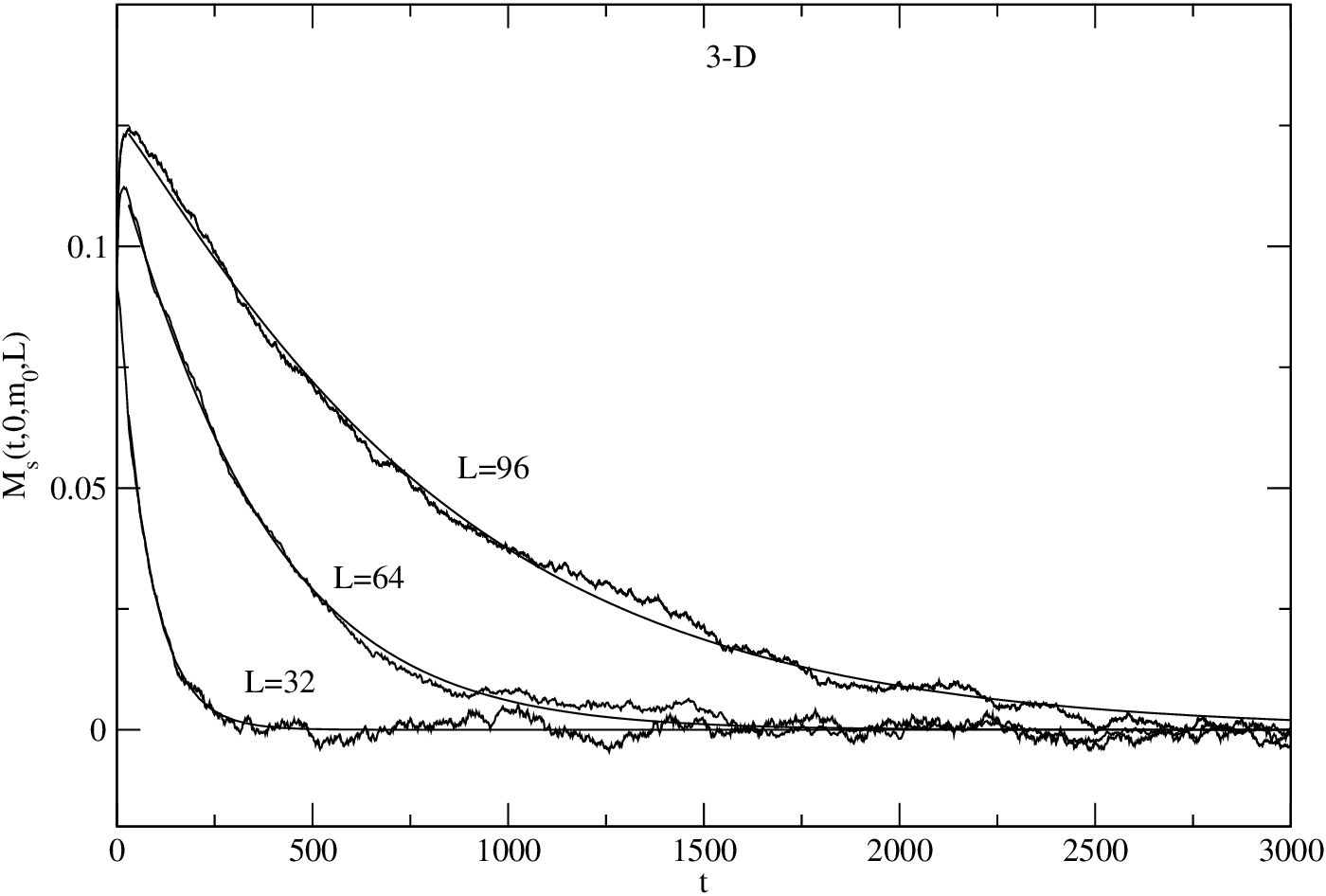}
\caption{}
\label{ttt2}
\end{figure}

\begin{figure}
\centering
\includegraphics[angle=0,width=10truecm]{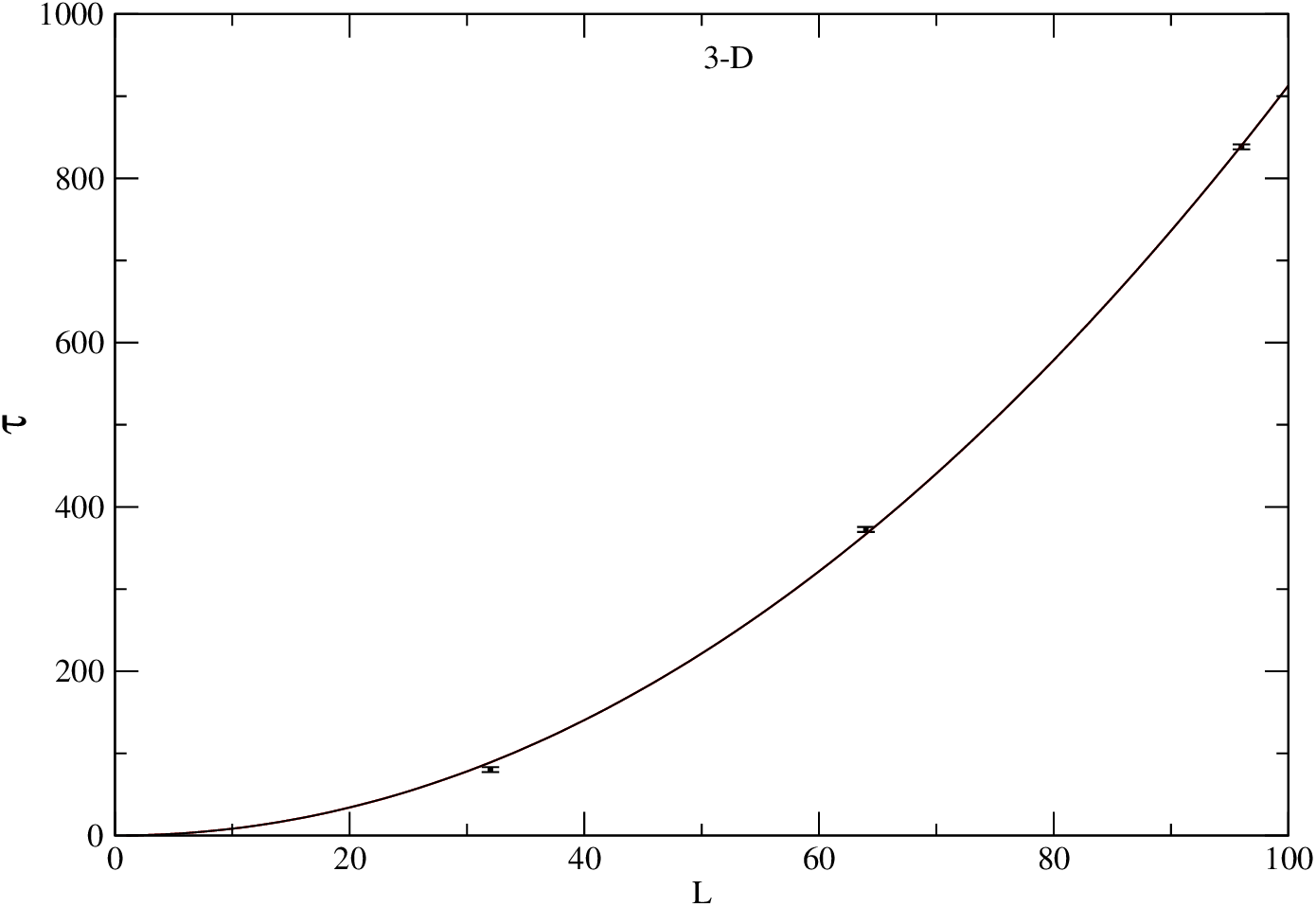}
\caption{}
\label{tttt3}
\end{figure}

\begin{figure}
\centering
\includegraphics[angle=0,width=10truecm]{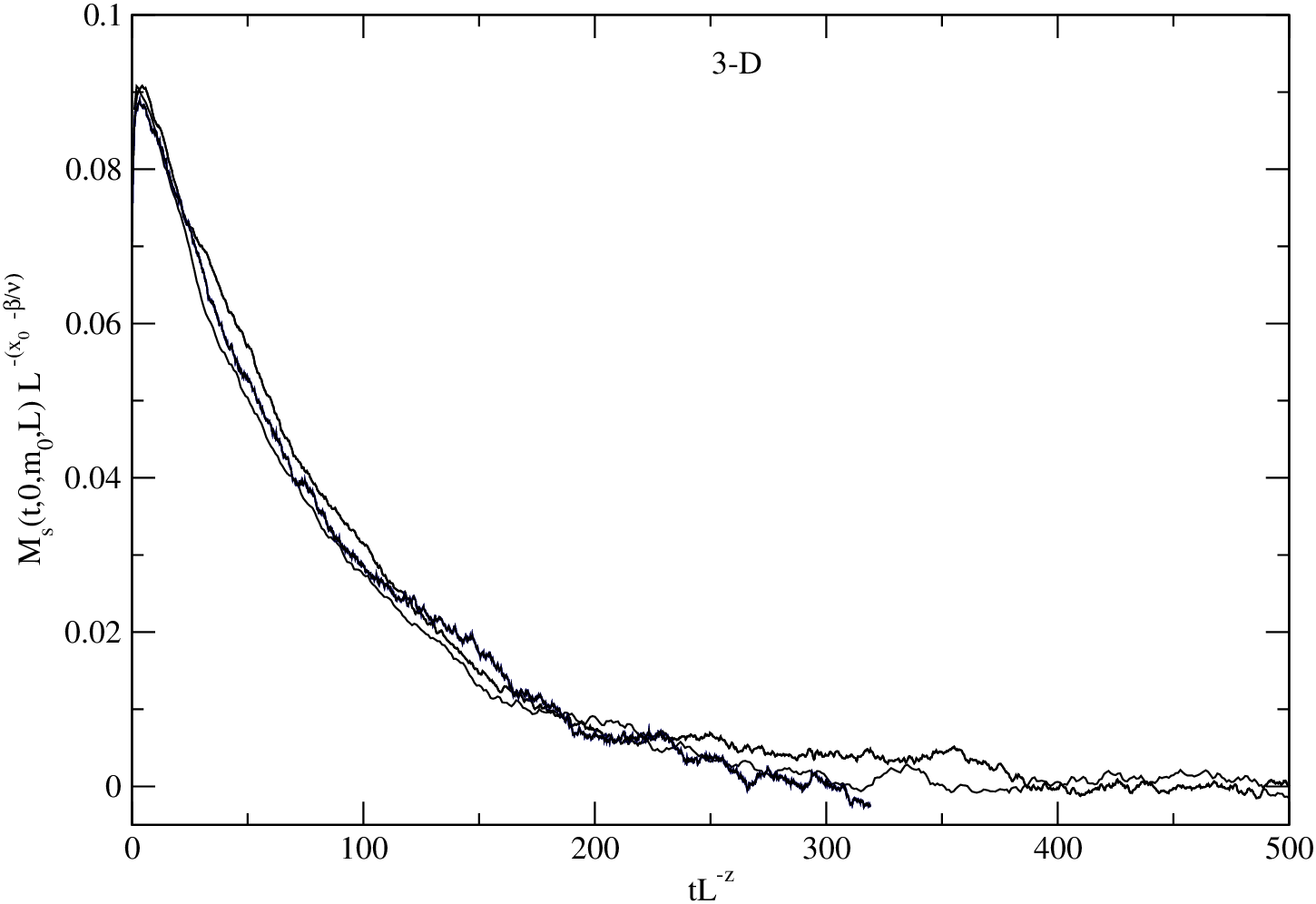}
\caption{}
\label{tttt4}
\end{figure}

\begin{figure}
\centering
\includegraphics[angle=0,width=10truecm]{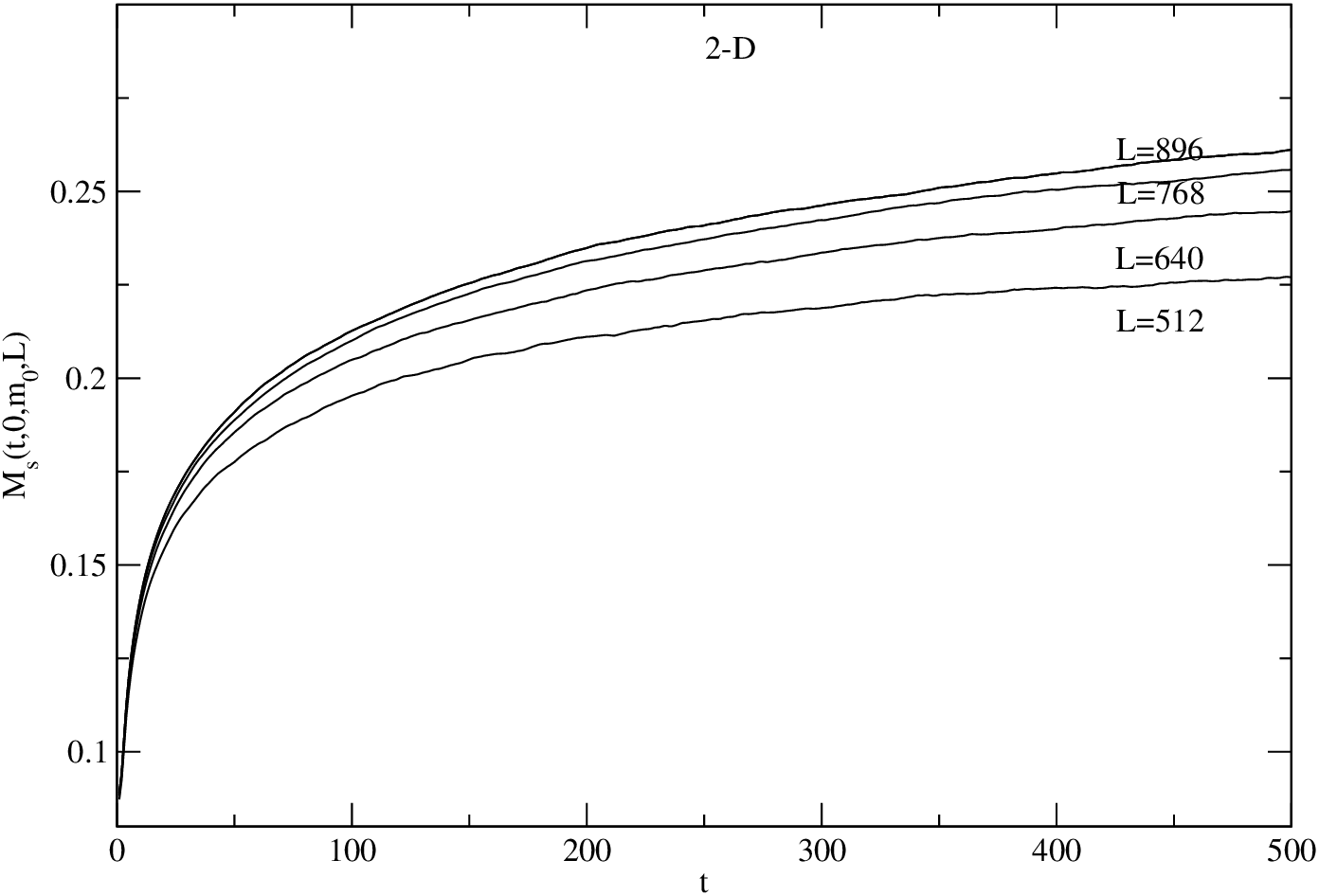}
\caption{}
\label{tttt5}
\end{figure}

\begin{figure}
\centering
\includegraphics[angle=0,width=10truecm]{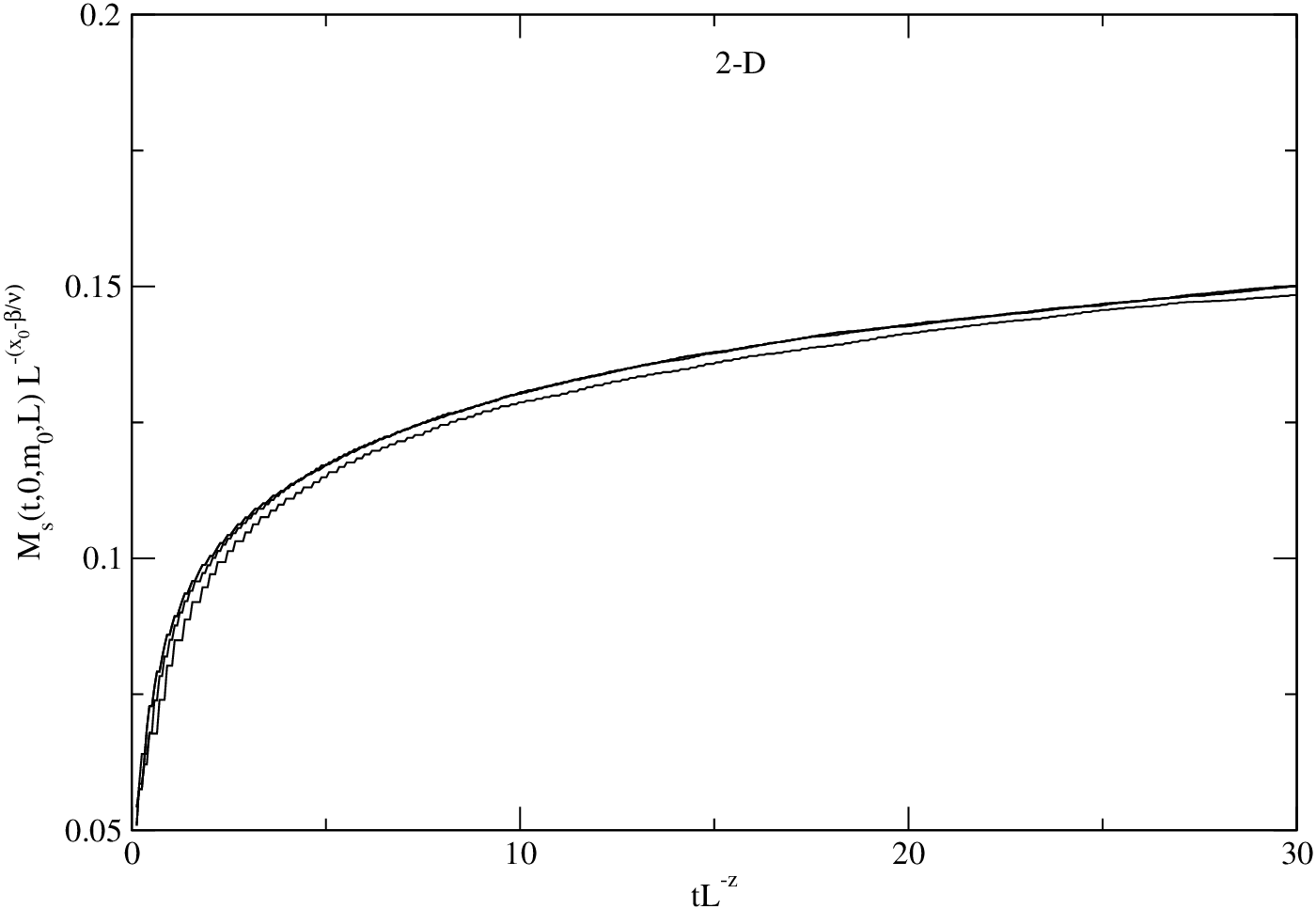}
\caption{}
\label{tttt6}
\end{figure}

\end{document}